# Multiparty Mediated Quantum Secret Sharing Protocol

Chia-Wei Tsai[1, a], Chun-Wei Yang[2, b], and Jason Lin[3, c]

[1]Department of Computer Science and Information Engineering, National Taitung University, No. 369, Sec. 2, University Rd., Taitung 95092, Taiwan

[2]Center for General Education, China Medical University, No. 100, Sec. 1, Jingmao Rd., Beitun Dist., Taichung City 406040, Taiwan

[2]Master Program for Digital Health Innovation, College of Humanities and Sciences, China Medical University, No. 100, Sec. 1, Jingmao Rd., Beitun Dist., Taichung City 406040, Taiwan

[3]Department of Computer Science and Engineering, National Chung Hsing University, No. 145, Xingda Rd., South District,

Taichung 40227, Taiwan

[a]cwtsai@nttu.edu.tw

[b]cwyang@mail.cmu.edu.tw

[c]jasonlin@nchu.edu.tw

## Abstract

This study proposes a multiparty mediated quantum secret sharing (MQSS) protocol that allows $n$ restricted quantum users to share a secret via the assistance of a dishonest third-party (TP) with full quantum capabilities. Under the premise that a restricted quantum user can only perform the Hadamard transformation and the Z-basis measurement, the proposed MQSS protocol has addressed two common challenges in the existing semi-quantum secret sharing protocols: (1) the dealer must have full quantum capability, and (2) the classical users must equip with the wavelength quantum filter and the photon number splitters (PNS) to detect the Trojan horse attacks. The security analysis has also delivered proof to show that the proposed MQSS protocol can avoid the collective attack, the collusion attack, and the Trojan horse attacks. In addition, the proposed MQSS protocol is more efficient than the existing SQSS protocols due to the restricted quantum users can only equip with two quantum operations, and the qubits are transmitted within a shorter distance.

## 1. Introduction

Secret sharing is an important research branch in conventional cryptography. The fundamental concept is that a dealer splits his/her secret message into several secret shadows, and then distributes them to multiple agents, respectively. Only when a sufficient number of agents cooperate using their secret shadows can the dealer's secret be recovered. In 1979, Shamir [1] proposed the first secret sharing protocol based on the idea that a polynomial of degree $n-1$ requires $n$ points to fit. Some other classical secret sharing (CSS) protocols have been proposed [2,3] using different mathematical properties such as geometry plane [2] and Chinese remainder theorem [3]. However, the security of these CSS protocols lies on the computational complexity of some mathematical problems using classical computer systems. Therefore, they are not suitable for the computing scenario using quantum mechanics. In 1999, Hillery et al. [4] proposed the first quantum secret sharing (QSS) protocol using Greenberger–Horne–Zeilinger (GHZ) states. After that, many QSS protocols [5-46] have been proposed based on different properties of quantum states. These QSS protocols always assume that both the dealer and agents have been equipped with complete quantum devices (e.g., quantum memory, quantum measurement device, and the generator for various quantum states). It is relatively unrealistic since some quantum devices are of high cost and difficult to implement. Hence, it is an interesting research issue in QSS to discuss whether all participants should have full quantum capabilities to execute the protocol. To address this issue, Boyer et al. [47] proposed a novel environment called semi-quantum environment with two kinds of users: the quantum users and the classical users. The quantum user has the complete quantum capabilities, but the classical user only equips with the limited quantum devices. According to the classical user's quantum capabilities, the semi-quantum environments can be classified into four types of semi-quantum environment which have been summarized in **Table 1**.

Table 1. The three kinds of semi-quantum environments

| Environment | Capabilities of classical user |
|---|---|
| Measure–Resend | (1) generate Z-basis qubits <br> (2) perform Z-basis measurement <br> (3) reflect photons without disturbance |
| Randomization-Based | (1) perform Z-basis measurement <br> (2) reorder photons by using different delay lines <br> (3) reflect photons without disturbance |
| Measurement-Free | (1) generate Z-basis qubits <br> (2) reorder photons by using different delay lines <br> (3) reflect photons without disturbance |

The first semi-quantum key distribution (SQKD) protocol was also proposed by Boyer et al. [47] in 2007. After that, various cryptographic protocols have been proposed under the semi-quantum environment. Among these semi-quantum cryptographic protocols, the semi-quantum secret sharing (SQSS) [48-62] is considered as one of the important research branches. In 2011, Li et al. [48] proposed the first SQSS protocol that allows a quantum dealer Alice to share her secrets with two classical agents, Bob, and Charlie, by using GHZ-type states. For three-particle quantum states, a more recent work was proposed by Tsai et al. [62] using W states. To expand the three-party idea into a more generalized scenario, Li et al. [50] further proposed a multiparty SQSS protocol using the product states. Their protocol was later improved by Yang and Hwang [51]. On the other hand, Xie et al. [52] proposed another multiparty SQSS protocol let the dealer can share his/her secret with the agents without additionally transmitting the additional ciphertext of the dealer's secret to the agents. For less particle entangled states, Wang et al. [49] and Gao et al. [50] proposed their multiparty SQSS scheme with two-particle entangled states and Bell states, respectively. Yi and Tong [60] proposed a more recent work using a unique two-particle entangled state to design their SQSS protocol. In 2018, Li et al. [55] proposed an SQSS scheme with limited quantum resources that allows all restricted quantum participants to have no quantum measurement devices. However, Tsai et al. [63] had pointed out that Li et al.'s protocol existed a double-CNOT attack. So far, these SQSS protocols have the following two restrictions: (1) the dealer that shares the secret cannot be any arbitrary classical quantum participant, (2) all classical quantum participants must equip with additional quantum devices such as the photon number splitters and the optical wavelength filter to avoid the Trojan horse attacks, which apparently violates the original assumption of the semi-quantum environment.

To overcome the above-mentioned two constraints, this study adopts the concept of mediated quantum communication protocol [64] to design the first multiparty mediated quantum secret sharing (MQSS) protocol with the $(n, n)$ threshold in which a restricted quantum dealer can share his/her secrets with the other restricted quantum agents via the help of a dishonest third-party (TP). It is worth to note that a restricted quantum participant indicates that the participant only owns limited quantum capabilities. However, the only difference between the restricted quantum user in a mediated quantum communication protocol and the classical user in a semi-quantum protocol is the former can equip with the unitary operators of a single qubit. The property of entanglement between the GHZ state and the Hadamard operation along with the one-way qubit transmission are used to complete the design of the proposed MQSS protocol. For the security aspect, our protocol is congenitally free from the Trojan Horse attacks due to the use of one-way qubit transmission. The design of our protocol process can also prevent both the collective attack and the collusion attack. For the efficiency aspect, both the restricted quantum dealer and the restricted quantum agents only require two quantum capabilities: (1) perform the Hadamard operator on a single qubit; (2) perform measurement on any single qubit with Z-basis $\{|0\rangle, |1\rangle\}$. Hence, the proposed MQSS protocol is considered to be more efficient than the exiting SQSS protocols in terms of the quantum capabilities of the participants and the distance of the qubit transmission. In addition, it is noteworthy that the proposed protocol also conforms to the quantum network infrastructure. Since the cost of quantum devices is unrealistic, a server with full quantum capabilities (say TP) that assists the edge/client devices with limited quantum capabilities to accomplish the communication tasks is a typical solution to reduce the construction cost of the network.

The rest of this paper is organized as follows: Section 2 describes the property of entanglement used to design the MQSS protocol, and the process of the MQSS are presented in Section 3. The security analysis and comparison are given in Section 4. Finally, a brief conclusion and some directions for future studies are provided in Section 5.

## 2. Entanglement property between GHZ state and Hadamard operation

Before describing the step-by-step procedure of the proposed MQSS scheme, this study first introduces the property between GHZ state and Hadamard operation that used to design the protocol.

The general form of the GHZ state and the Hadamard operation is expressed in **Eq. (1)** and **Eq. (2)**, respectively, as follows:

$$|\Psi_{(x_1x_2\cdots x_n,b)}\rangle_{12\cdots n} = \frac{1}{\sqrt{2}}(|x_1x_2\cdots x_n\rangle + (-1)^b|\overline{x_1x_2\cdots x_n}\rangle)_{12\cdots n} \quad (1),$$

where $x_i \in \{0,1\}$ and $b \in \{0,1\}$.

$$H = \frac{1}{\sqrt{2}}[(|0\rangle + |1\rangle)\langle 0| + (|0\rangle - |1\rangle)\langle 1|] \quad (2).$$

Assume that each particle of a GHZ state can perform either identity operator or Hadamard operation. With the above operations, the GHZ state will project to one of the following four properties.

**Property 1:** Since an identity operator will not change the state of a vector, it is obvious that the GHZ state will remain invariant if all particles have been multiplied with the identity operators.

**Property 2:** If all particles of the GHZ state have been passed through the logic gate of the Hadamard operation, the state will convert to the GHZ-like state as shown in the following expression:

$$|G_{(x_1x_2\cdots x_n,b)}\rangle_{12\cdots n} = \frac{1}{\sqrt{2^{n-1}}}\sum_{1}^{2^{n-1}}[(-1)^\delta|k_1k_2\ldots k_n\rangle]_{num(1)} \quad (3),$$

where $k_i \in \{0,1\}$, $\delta = \sum_{i=1}^{n}k_ix_i$, and $num(1)$ denotes the total number of "1" is even or odd in each vector $|k_1k_2\ldots k_n\rangle$. If $b = 0$, $num(1)$ will be even; otherwise, $num(1)$ will be odd. Here, we take the 4-particle GHZ state $|\Psi_{(0011,0)}\rangle_{1234} = \frac{1}{\sqrt{2}}(|0011\rangle + |1100\rangle)_{1234}$ and $|\Psi_{(0011,1)}\rangle_{1234} = \frac{1}{\sqrt{2}}(|0011\rangle - |1100\rangle)_{1234}$ as the example to explain the properties between GHZ state and Hadamard operation. After performing Hadamard operations on all qubits, the system states are changed as follows:

$$H^{\otimes 4}|\Psi_{(0011,0)}\rangle_{1234} = |G_{(0011,0)}\rangle_{1234} = \frac{1}{2\sqrt{2}}\begin{pmatrix}|0000\rangle + |0011\rangle - |0101\rangle - |0110\rangle - \\ |1001\rangle - |1010\rangle + |1100\rangle + |1111\rangle\end{pmatrix}_{1234} \quad (4),$$

$$H^{\otimes 4}|\Psi_{(0011,1)}\rangle_{1234} = |G_{(0011,1)}\rangle_{1234} = \frac{1}{2\sqrt{2}}\begin{pmatrix}-|0001\rangle - |0010\rangle + |0100\rangle + |0111\rangle + \\ |1000\rangle + |1011\rangle - |1101\rangle - |1110\rangle\end{pmatrix}_{1234} \quad (5).$$

From **Eq. (3)** to **Eq. (5)**, we found that the result of $\oplus_{i=1}^{n} mr_{Z-basis}^{i} = 0$ if $b$ of the initial GHZ state is 0, and the result of $\oplus_{i=1}^{n} mr_{Z-basis}^{i} = 1$ if $b = 1$, where $\oplus$ indicates the exclusive-OR operation and $mr_{Z-basis}^{i}$ is the measurement result of the $i$-th particle using Z-basis.

**Property 3:** If any $n - 1$ particles of a GHZ state have been performed with the Hadamard operation, the GHZ state will convert to the following state.

$$|\Phi_{(x_1x_2\cdots x_n,b)}\rangle_{12\cdots n} = \frac{1}{\sqrt{2^{n-1}}}\sum_{K=0}^{2^{n-1}-1}(-1)^\delta|k_1k_2\ldots k_{n-1}\rangle \otimes \frac{1}{\sqrt{2}}\left(|x_{p_j^I}\rangle + (-1)^{\Delta+b}|\overline{x_{p_j^I}}\rangle\right) \quad (6),$$

where $k_1k_2\ldots k_m$ is the binary form of the state $K$ (i.e., $k_i \in \{0,1\}$), two subscripts $p_i^H$ and $p_j^I$ are defined as $p_i^H \in \{1,2,\ldots,n\}$ for $\forall i \in \{1,2,\ldots,n-1\}$ and $p_j^I \in \{1,2,\ldots,n\}$ for $\forall j \in \{1\}$, and $p_i^H$ ($p_j^I$) is the index for each position of particle that perform the Hadamard (Identity) operation. Then, $\delta$ and $\Delta$ are defined as follows:

$$\delta = \begin{cases} \sum_{i=1}^{n-1} k_i x_{p_i^H}, & \text{if } x_{p_j^I} = 0 \\ \sum_{i=1}^{n-1} k_i \overline{x_{p_i^H}}, & \text{if } x_{p_j^I} = 1 \end{cases} \quad (7),$$

and

$$\Delta = \begin{cases} 0, & \text{if hamming weight of } (k_1k_2\ldots k_{n-1}) \text{ is even} \\ 1, & \text{if hamming weight of } (k_1k_2\ldots k_{n-1}) \text{ is odd} \end{cases} \quad (8).$$

Here, we take the 4-particle GHZ state (i.e., $|\Psi_{(0010,0)}\rangle_{1234} = \frac{1}{\sqrt{2}}(|0010\rangle + |1101\rangle)_{1234}$ and $|\Psi_{(0001,0)}\rangle_{1234} = \frac{1}{\sqrt{2}}(|0001\rangle + |1110\rangle)_{1234}$) as an example. Suppose that the Hadamard operations are performed on the 1st, 2nd and 3rd particles of GHZ states. The entanglement system after the transformations will change to the following states:

$$H^{\otimes 3} \otimes I |\Psi_{(0010,0)}\rangle_{1234} = \frac{1}{2\sqrt{2}} \begin{pmatrix} |000\rangle_{123}\otimes|+\rangle_4 - |001\rangle_{123}\otimes|-\rangle_4 + |010\rangle_{123}\otimes|-\rangle_4 \\ -|011\rangle_{123}\otimes|+\rangle_4 + |100\rangle_{123}\otimes|-\rangle_4 - |101\rangle_{123}\otimes|+\rangle_4 \\ +|110\rangle_{123}\otimes|+\rangle_4 - |111\rangle_{123}\otimes|-\rangle_4 \end{pmatrix}_{1234}$$

(9),

and

$$H^{\otimes 3} \otimes I |\Psi_{(0001,0)}\rangle_{1234} = \frac{1}{2\sqrt{2}} \begin{pmatrix} |000\rangle_{123}\otimes|+\rangle_4 - |001\rangle_{123}\otimes|-\rangle_4 - |010\rangle_{123}\otimes|-\rangle_4 \\ +|011\rangle_{123}\otimes|+\rangle_4 - |100\rangle_{123}\otimes|-\rangle_4 + |101\rangle_{123}\otimes|+\rangle_4 \\ +|110\rangle_{123}\otimes|+\rangle_4 - |111\rangle_{123}\otimes|-\rangle_4 \end{pmatrix}_{1234}$$

(10),

where $|+\rangle = \frac{1}{\sqrt{2}}(|0\rangle + |1\rangle)$ and $|-\rangle = \frac{1}{\sqrt{2}}(|0\rangle - |1\rangle)$.

From **Eq. (6)**, **Eq. (9)**, and **Eq. (10)**, we found that the state of the particle without Hadamard operation will convert to X-basis ($|+\rangle, |-\rangle$), and the other particles have lost entanglement correlation if they use Z-basis to measure the particles.

**Property 4:** If $m$ particles of GHZ state where $1 \leq m \leq n-2$ have been performed with the Hadamard operation, the GHZ state will result in the following state.

$$|\Phi_{(x_1x_2\cdots x_n,b)}\rangle_{12\cdots n} = \frac{1}{\sqrt{2^m}}\sum_{K=0}^{2^m-1}(-1)^\delta|k_1k_2\ldots k_m\rangle\otimes\frac{1}{\sqrt{2}}\left(|x_{p_1^I}x_{p_2^I}\ldots x_{p_{n-m}^I}\rangle + (-1)^{\Delta+b}|\overline{x_{p_1^I}x_{p_2^I}\ldots x_{p_{n-m}^I}}\rangle\right)$$

(11),

where $k_1k_2\ldots k_m$ is the binary form of the state $K$ (i.e., $k_i \in \{0,1\}$), the two subscripts $p_i^H$ and $p_j^I$ are also defined as $p_i^H \in \{1,2,\ldots,n\}$ for $\forall i \in \{1,2,\ldots,m\}$ and $p_j^I \in \{1,2,\ldots,n\}$ for $\forall j \in \{1,2,\ldots,n-m\}$, $p_i^H$ ($p_j^I$) also represents the index for each position of particle that perform the Hadamard (Identity) operation, and $L$ is the decimal form of $x_{p_1^I}x_{p_2^I}\ldots x_{p_{n-m}^I}$. Then, $\delta$ and $\Delta$ are defined as **Eq. (12)** and **(13)**.

$$\delta = \begin{cases} \sum_{i=1}^{n-1} k_i x_{p_i^H}, & \text{if } L < 2^{n-m-1} \\ \sum_{i=1}^{n-1} k_i \overline{x_{p_i^H}}, & \text{if } L \geq 2^{n-m-1} \end{cases} \quad (12),$$

and

$$\Delta = \begin{cases} 0, & \text{if hamming weight of } (k_1k_2\ldots k_m) \text{ is even} \\ 1, & \text{if hamming weight of } (k_1k_2\ldots k_m) \text{ is odd} \end{cases} \quad (13).$$

Here, we also take the 4-particle GHZ state $|\Psi_{(0001,0)}\rangle_{1234} = \frac{1}{\sqrt{2}}(|0010\rangle + |1101\rangle)_{1234}$ and $|\Psi_{(0000,0)}\rangle_{1234} = \frac{1}{\sqrt{2}}(|0000\rangle + |1111\rangle)_{1234})$ as an example to demonstrate the property. Suppose that the Hadamard operations are perform on either the 1st and 2nd particles or only the 1st particle of the GHZ states. After the transformation, the state of the original entanglement system will change to the followings:

$$H\otimes H\otimes I\otimes I|\Psi_{(0001,0)}\rangle_{1234} = \frac{1}{2}\begin{pmatrix}|00\rangle_{12}\otimes\frac{1}{\sqrt{2}}(|01\rangle+|10\rangle)_{34} - |01\rangle_{12}\otimes\frac{1}{\sqrt{2}}(|01\rangle-|10\rangle)_{34} - \\ |10\rangle_{12}\otimes\frac{1}{\sqrt{2}}(|01\rangle-|10\rangle)_{34} + |11\rangle_{12}\otimes\frac{1}{\sqrt{2}}(|01\rangle+|10\rangle)_{34}\end{pmatrix}$$

(14),

$$H\otimes H\otimes I\otimes I|\Psi_{(0000,0)}\rangle_{1234} = \frac{1}{2}\begin{pmatrix}|00\rangle_{12}\otimes\frac{1}{\sqrt{2}}(|00\rangle+|11\rangle)_{34} + |01\rangle_{12}\otimes\frac{1}{\sqrt{2}}(|00\rangle-|11\rangle)_{34} + \\ |10\rangle_{12}\otimes\frac{1}{\sqrt{2}}(|00\rangle-|11\rangle)_{34} + |11\rangle_{12}\otimes\frac{1}{\sqrt{2}}(|00\rangle+|11\rangle)_{34}\end{pmatrix}$$

(15),

$$H\otimes I^{\otimes 3}|\Psi_{(0001,0)}\rangle_{1234} = \frac{1}{\sqrt{2}}\left(|0\rangle_1\otimes\frac{1}{\sqrt{2}}(|010\rangle+|101\rangle)_{234} + |1\rangle_1\otimes\frac{1}{\sqrt{2}}(|010\rangle-|101\rangle)_{234}\right)$$

(16),

and

$$H\otimes I^{\otimes 3}|\Psi_{(0000,0)}\rangle_{1234} = \frac{1}{\sqrt{2}}\left(|0\rangle_1\otimes\frac{1}{\sqrt{2}}(|000\rangle+|111\rangle)_{234} + |1\rangle_1\otimes\frac{1}{\sqrt{2}}(|000\rangle-|111\rangle)_{234}\right)$$

(17).

From **Eq. (11)**, and **Eq. (14)** to **Eq. (17)**, we found that measurement result of those particles without performing the Hadamard operations will be a maximally entangled state. That is, Bell state if the number of these particles is equal to 2, or otherwise GHZ state if the number of these particles is greater than or equal to 3. Note that the phase of the remaining entanglement state is based on the Hamming weight of measurement results for particles performed by Hadamard operations in Z-basis. Taking **Eq. (16)** and **(17)** as the examples, compared with the phase of the initial GHZ state, the remaining entanglement state (2nd, 3rd and 4th particles) will be changed from positive to negative when the Hamming weight of the measurement result for the 1st particle is odd; otherwise, the phase of the remaining entanglement state will remain invariant.

### 3. Proposed MQSS Protocol

In this section, some initial assumptions on the quantum capabilities are given to better illustrate the procedure of the proposed MQSS protocol with the $(n, n)$ threshold. The restricted quantum participants (i.e., the dealer and the agents) only has two quantum capabilities: (1) the Hadamard operation, and (2) the Z-basis measurement ($|0\rangle, |1\rangle$); TP can generate the GHZ states and store these states using the quantum memory. To be more realistic in practice, the security level of TP is set to allow dishonesty. That is, TP may perform any possible attacks such as colluding with other insiders to learn the information of the dealer's secret and/or the other agents' secret shadows. For qubit transmission, quantum channels only exist between TP and each restricted quantum participant with a predefined noise rate $\varepsilon$. All the restricted quantum participants have mutually owned an authenticated classical channel between each other in which the messages can be eavesdropped but cannot be modified. For the procedure of the protocol, this study will first deliver its general form and then demonstrates a five-party communication scenario as an example.

Suppose that a dealer Alice wants to share a $m$-bit secret with $n$ agents (i.e., Bob$_1$, Bob$_2$, …, Bob$_n$) via the help of a third-party server TP. The step-by-step process of the proposed MQSS protocol are depicted as follows.

**Step 1.** TP generates any GHZ state $|\Psi_{(x_1 x_2 \cdots x_n, b)}\rangle_{12 \cdots n+1} = \frac{1}{\sqrt{2}}(|x_1 x_2 \cdots x_n x_{n+1}\rangle + (-1)^b |\overline{x_1 x_2 \cdots x_n x_{n+1}}\rangle)_{12 \cdots n+1}$, and sends the 1st particle to Alice and keeps the remaining particles in quantum memory.

**Step 2.** After Alice receiving the qubit, she randomly selects **Check** or **Share** mode. If Alice selects **Share** modes, she will perform Hadamard operation on the received qubit and then measure it with Z-basis; otherwise, she will directly measure the received qubit in Z-basis. Alice stores the measurement result $mr_A^i$ where $i$ indicates the $i$-th transmission. Then, she returns the acknowledgement (i.e., ACK) signal to TP. According to the properties of **Eq. (11)**, we know that the phase of the remaining particles is based on the measurement result $mr_A^i$. The phase will be flipped (i.e., $-$ to $+$ or $+$ to $-$) if $mr_A^i = 1$; otherwise, the phase will remain invariant.

**Step 3.** After receiving the ACK signal from Alice, TP will then send the remaining particles to each agent, respectively.

**Step 4.** Each agent also randomly selects **Check** or **Share** mode. If the agent selects **Share** modes, he will perform the Hadamard operation on the received qubit and then measure it with Z-basis. Otherwise, he will directly perform the Z-basis measurement on the received qubit. After the measurement, all agents will store the results $mr_{B1}^i$, $mr_{B2}^i$,…, $mr_{Bn}^i$, respectively. Repeat **Step 1** to **Step 4** several times with a suggestion of at least $m \times 2^{n+1}$ times due to the following reasons.

- In each qubit transmission between TP and the restricted quantum participant, about half of the qubits will be used for eavesdropping check, so only half of the measurement results will become candidates of a potential shared secret key.
- The measurement results possessed by each restricted quantum participant will have about half of them either discard or used for eavesdropping check. Therefore, the qubit efficiency used for the shared one-bit secret will be calculated as $\frac{1}{2} \times \frac{1}{2^n} = \frac{1}{2^{n+1}}$.

**Step 5.** TP announces all initial states of those GHZ states generated in Step 1. Then, Alice requests the agents to disclose the information of their mode selections in Step 4. Table 2 summarizes the three possible cases that will happen according to the modes selected by Alice and the three agents.

From the selected modes in **Step 4**, Alice will ask the agents to announce their corresponding measurement results in **Case 2** and **Case 3** of **Table 2** to execute the public discussion for security checking of TP and eavesdroppers. If the error rate exceeds the noise rate $\varepsilon$ of the quantum channel, the participants will abort the session and restart the protocol. Otherwise, the participants will share the raw key bits as depicted in **Case 1** of **Table 2**. Here, Alice and the agents should share $2m$-bit raw key bits because the participants repeat **Step 1** to **Step 4** $m \times 2^{n+1}$ times and each participant selects **Share** mode with the probability of 50%.

**Step 6.** Alice selects $m$-bit of raw secret key as the check bits $CK_A = (ck_A^1, ck_A^2, ..., ck_A^m)$, and then she requests all agents to announce their check bit values (i.e., $(ck_{B1}^1, ck_{B1}^2, ..., ck_{B1}^m)$, $(ck_{B2}^1, ck_{B2}^2, ..., ck_{B2}^m)$,…, and $(ck_{Bn}^1, ck_{Bn}^2, ..., ck_{Bn}^m)$). Alice checks the values depending on whether they conform to the equation $ck_A^j = \bigoplus_{x=1}^{n} ck_{Bx}^j$ for $1 \leq j \leq$

*m*. If the error rate exceeds the preset threshold $\varepsilon$, the current session will be aborted for relaunching the protocol.

**Step 7.** Alice and the agents will take the remaining *m*-bit raw shadow bits as the shared secret key: $SK_A = \{sk_A^1, sk_A^2, \cdots, sk_A^m\}$, $SK_{B1} = \{s_{B1}^1, s_{B1}^2, \cdots, s_{B1}^m\}$, $SK_{B2} = \{s_{B2}^1, s_{B2}^2, \cdots, s_{B2}^m\}, \ldots,$ and $SK_{Bn} = \{s_{Bn}^1, s_{Bn}^2, \cdots, s_{Bn}^m\}$, respectively. Then, Alice announces her secret $S$ via the result of $S \oplus SK_A$.

**Step 8.** Only when all agents collaborate with each other can they recover Alice's shared secret bits (i.e., $S \oplus SK_A \oplus_{x=1}^n SK_{Bx} = S$).

**Table 2.** the possible cases among Alice's and the agents' modes

| Case | Description | Strategy |
|---|---|---|
| 1 | Alice and all agents select **Share** mode in which they all perform Hadamard operations on their corresponding qubits. | Alice and all agents first collect each corresponding measurement result as the raw keys respectively. Then, they take a portion of the raw key bits to perform a security check of TP and eavesdroppers in **Step 6** of the proposed protocol. |
| 2 | All participants select **Check** mode in which they all perform identity operators on their corresponding qubits. | The corresponding measurement results will be used to perform a security check of TP and eavesdroppers in **Step 5** of the proposed protocol. |
| 3 | The number of participants that select **Check** mode is greater than or equal to 2. | The participants who select **Check** mode will take their measurement results to perform a security check of TP and eavesdroppers in **Step 5** of the proposed protocol, while the other participants who select **Share** mode will discard their measurement results. |

This study takes a five-party communication scenario as an example to further demonstrate the idea of the proposed MQSS protocol. In the example scenario, assume that the dealer Alice wants to share a *m*-bit secret with the three agents: Bob, Charlie, and David, via the help of a third-party server TP. The process of the proposed MQSS protocol is depicted as follows (also shown in **Figure 1**).

**Step 1.** TP generates any GHZ state $|\Psi_{(x_1x_2x_3x_4,b)}\rangle = \frac{1}{\sqrt{2}}(|x_1x_2x_3x_4\rangle + (-1)^b|\overline{x_1x_2x_3x_4}\rangle)_{1234}$, and then sends the 1st particle to Alice and keeps the remaining particles in quantum memory.

**Step 2.** After Alice receiving the qubit, she randomly selects **Check** or **Share** mode. If Alice selects **Share** modes, she will perform Hadamard operation on the received qubit and then measure it with Z-basis; otherwise, she will directly measure the received qubit in Z-basis. Alice stores the measurement result $mr_A^i$ and returns the ACK signal to TP.

**Step 3.** After receiving the ACK signal from Alice, TP will then send the 2nd, 3rd, and 4th particles to Bob, Charlie, and David, respectively.

**Step 4.** Each agent also randomly selects **Check** or **Share** mode, performs the corresponding operations, and the Z-basis measurement on each of the received qubits. Here, Bob, Charlie, and David store the measurement results $mr_B^i$, $mr_C^i$, and $mr_D^i$, respectively. Alice and the agents will then repeat **Step 1** to **Step 4** $32m$ ($m \times 2^{4+1}$) times.

**Step 5.** TP announces all initial states of those GHZ states generated in Step 1. Then, Alice requests the agents to disclose the information of their mode selections in **Step 4.**

From the selected modes in **Step 4**, Alice executes the public discussions with all the agents for security checking of TP and eavesdroppers in **Case 2** and **Case 3** of **Table 2**. If the error rate exceeds the preset noise rate $\varepsilon$ of the quantum channel, they will abort the protocol and restart it again. Otherwise, the participants will obtain the raw key bits as depicted in **Case 1** of **Table 2**.

**Step 6.** Alice selects $m$-bit of raw secret key as the check bits $CK_A = (ck_A^1, ck_A^2, \ldots, ck_A^m)$, and then she requests all agents to announce their values of check bits. That is, $(ck_B^1, ck_B^2, \ldots, ck_B^m)$, $(ck_C^1, ck_C^2, \ldots, ck_C^m)$, and $(ck_D^1, ck_D^2, \ldots, ck_D^m)$). Alice checks the values depending on whether they conform to the equation $ck_A^j = ck_B^j \oplus ck_C^j \oplus ck_D^j$ for $1 \leq j \leq m$. If the error rate exceeds the preset threshold $\varepsilon$, the current session will be aborted and relaunch the protocol.

**Step 7.** Alice, Bob, Charlie, and David will take the remaining raw shadow bits as the shared secret key: $SK_A$ $SK_B$, $SK_C$, and $SK_D$, respectively. Then, Alice announces her secret $S$ via the result of $S \oplus SK_A$.

**Step 8.** Only when all three classical agents Bob, Charlie, and David collaborate together can they recover Alice's shared secret bit (i.e., $S \oplus SK_A \oplus SK_B \oplus SK_C \oplus SK_D = S$).

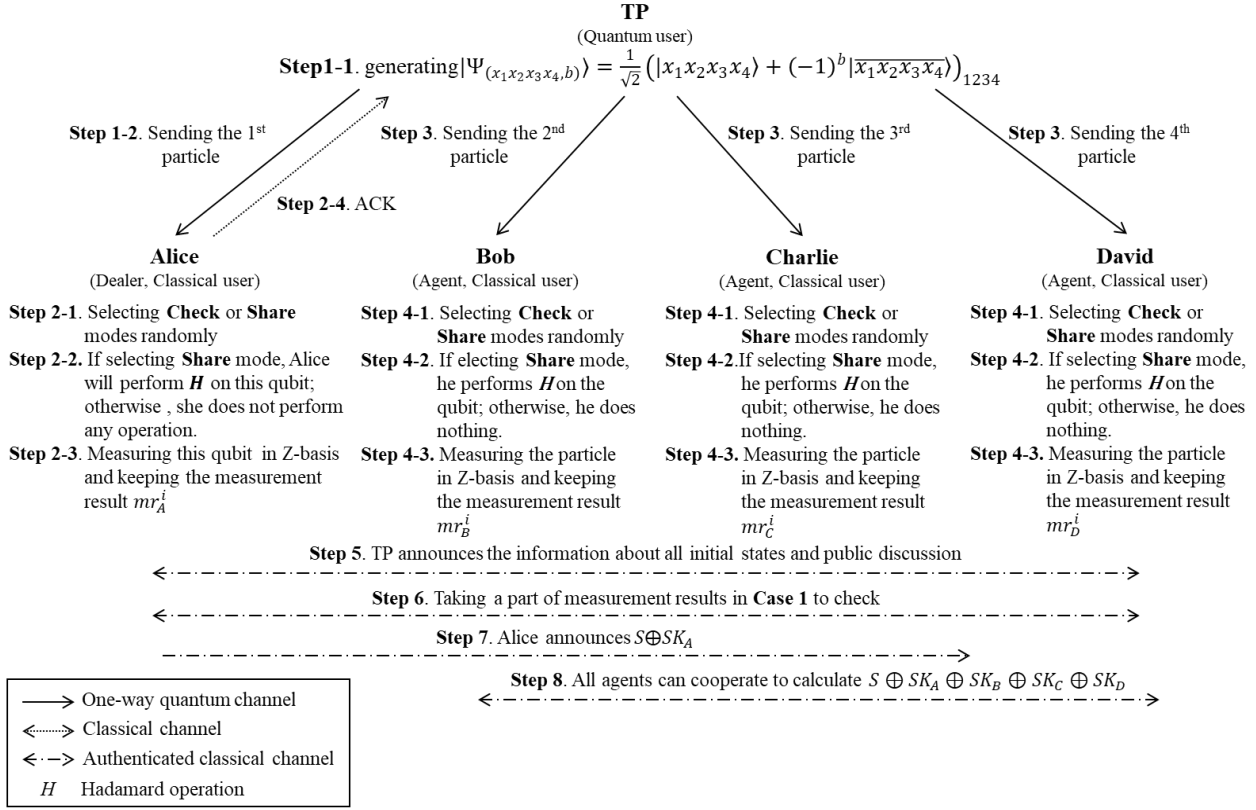

**Figure 1.** The process of the proposed MQSS protocol

## 4. Security Analysis

In this section, we analyze the security of the proposed MQSS protocol. For these quantum protocols, there are three attack modes including an individual attack, a collective attack, and a coherence attack [65,66]. The individual attack and the coherence attack have the most and the least restrictions, respectively. To date, no research work has pointed out that an attacker can obtain more advantages by using the coherence attack over the collective attack which is the most common attack in SQSS protocols [64,67-69]. Therefore, we take the proposed MQSS protocol with 5 participants as an example, give a complete analysis of the collective attack to prove the proposed MQSS protocol is robust under the attack intention of TP, and then the collusion attack is discussed to prove that the protocol is still secure if any insider has been compromised to collaborate with an attacker. Furthermore, an illustration of why the proposed MQSS protocol is congenitally free from the Trojan horse attacks will be given at the end of this section.

**4-1 Collective Attack**

In the proposed MQSS protocol, TP has many advantages over the other internal and external communicating parties to obtain the information of Alice's secret and/or the other agents' secret shadows. To prove that the proposed MQSS is robust under the collective attack, we must show TP's malicious behavior can be detected by the security check of the legitimate participants. Here, we take the five-party MQSS protocol as an example to analyze collective attack.

**Theorem 1**: TP performs the collective attack on the qubit that sent to Alice and the other agents. In order to do so, TP applies a unitary operation $U_e$ on the qubit that complies with the theorems of quantum mechanics. However, no unitary operation exists in the collective attack that allows TP to obtain the information of the participants' secret without being detected.

**Proof:** Before TP sends the photons of GHZ states to Alice and the other agents, TP performs a unitary operation $U_e$ on each GHZ state with an attached probe qubit $|E^i\rangle$ where $1 \leq i \leq 32m$. TP will keep the probe qubits in its quantum memory, and then measure the probe qubits with the corresponding basis to extract the information of Alice's secret and/or the agents' secret shadows after the completion of the protocol. For simplicity, we take one kind of GHZ state, $|\Psi^+\rangle_{1234} = \frac{1}{\sqrt{2}}(|0000\rangle + |1111\rangle)_{123}$, as an example to demonstrate the proof. Note that other GHZ states can apply the same approach to prove its robustness. According to the theorems of quantum mechanics, the entanglement system will convert to the following form after performing $U_e$ on the GHZ state with the probe qubit $|E^i\rangle$.

$$U_e|\Psi^+\rangle_{1234} \otimes |E^i\rangle = \alpha_0|0000\rangle|e_0^i\rangle + \alpha_1|0001\rangle|e_1^i\rangle + \alpha_2|0010\rangle|e_2^i\rangle + \alpha_3|0011\rangle|e_3^i\rangle + \alpha_4|0100\rangle|e_4^i\rangle + \alpha_5|0101\rangle|e_5^i\rangle + \alpha_6|0110\rangle|e_6^i\rangle + \alpha_7|0111\rangle|e_7^i\rangle + \alpha_8|1000\rangle|e_8^i\rangle + \alpha_9|1001\rangle|e_9^i\rangle + \alpha_{10}|1010\rangle|e_{10}^i\rangle + \alpha_{11}|1011\rangle|e_{11}^i\rangle + \alpha_{12}|1100\rangle|e_{12}^i\rangle + \alpha_{13}|1101\rangle|e_{13}^i\rangle + \alpha_{14}|1110\rangle|e_{14}^i\rangle + \alpha_{15}|1111\rangle|e_{15}^i\rangle \quad (18),$$

where $i$ indicates the $i$-th transmission and $\sum_{j=0}^{15}|\alpha_j|^2 = 1$. The state of the probe qubit $|e_j^i\rangle$ for $\forall j \in \{1,2,...,15\}$ can be distinguished by Eve due to $|e_j^x\rangle$ and $|e_j^y\rangle$ are orthogonal when $x \neq y$.

If TP attempts to pass the security check in **Case 2** of **Table 2**, it needs to adjust $U_e$ to make **Eq. (18)** conforms to the measurement results of GHZ state. Here, TP can set $\alpha_j = 0$ for $\forall j \in \{2,...,14\}$ to convert the system to the following state:

$$U_e|\Psi^+\rangle_{1234} \otimes |E^i\rangle = \alpha_0|0000\rangle|e_0^i\rangle + \alpha_{15}|1111\rangle|e_{15}^i\rangle \quad (19),$$

where $|\alpha_0|^2 + |\alpha_{15}|^2 = 1$.

According to the **Property 3** in **Section 2**, it shows that TP can also pass the security check in **Case 3** of **Table 2** when using the same settings of $U_e$. To further pass the security check in **Case 1** of **Table 2**, TP needs to adjust $U_e$ to let $mr_A^i \oplus mr_B^i \oplus mr_C^i \oplus mr_D^i = 0$. Given that there are two kinds of quantum state depending on Alice's measurement result when Alice selects **Share mode** and performs the corresponding operations in **Step 2**. That is, if $mr_A^i = 0$, the quantum state will become **Eq. (20)**; otherwise, the quantum state will become **Eq. (21)**.

$$\alpha_0|000\rangle|e_0^i\rangle + \alpha_{15}|111\rangle|e_{15}^i\rangle \quad (20),$$

$$\alpha_0|000\rangle|e_0^i\rangle - \alpha_{15}|111\rangle|e_{15}^i\rangle \quad (21).$$

After all agents perform the Hadamard operations, the states of **Eq. (20)** and **Eq. (21)** will become **Eq. (22)** and **Eq. (23)**, respectively.

$$\frac{1}{2\sqrt{2}}\Big(|000\rangle \otimes (\alpha_0|e_0^i\rangle + \alpha_{15}|e_{15}^i\rangle) + |001\rangle \otimes (\alpha_0|e_0^i\rangle - \alpha_{15}|e_{15}^i\rangle) + |010\rangle \otimes (\alpha_0|e_0^i\rangle - \alpha_{15}|e_{15}^i\rangle) + |011\rangle \otimes (\alpha_0|e_0^i\rangle + \alpha_{15}|e_{15}^i\rangle) + |100\rangle \otimes (\alpha_0|e_0^i\rangle - \alpha_{15}|e_{15}^i\rangle) + |101\rangle \otimes (\alpha_0|e_0^i\rangle + \alpha_{15}|e_{15}^i\rangle) + |110\rangle \otimes (\alpha_0|e_0^i\rangle + \alpha_{15}|e_{15}^i\rangle) + |111\rangle \otimes (\alpha_0|e_0^i\rangle - \alpha_{15}|e_{15}^i\rangle)\Big) \quad (22),$$

$$\frac{1}{2\sqrt{2}}\Big(|000\rangle \otimes (\alpha_0|e_0^i\rangle - \alpha_{15}|e_{15}^i\rangle) + |001\rangle \otimes (\alpha_0|e_0^i\rangle + \alpha_{15}|e_{15}^i\rangle) + |010\rangle \otimes (\alpha_0|e_0^i\rangle + \alpha_{15}|e_{15}^i\rangle) + |011\rangle \otimes (\alpha_0|e_0^i\rangle - \alpha_{15}|e_{15}^i\rangle) + |100\rangle \otimes (\alpha_0|e_0^i\rangle + \alpha_{15}|e_{15}^i\rangle) + |101\rangle \otimes (\alpha_0|e_0^i\rangle - \alpha_{15}|e_{15}^i\rangle) + |110\rangle \otimes (\alpha_0|e_0^i\rangle - \alpha_{15}|e_{15}^i\rangle) + |111\rangle \otimes (\alpha_0|e_0^i\rangle + \alpha_{15}|e_{15}^i\rangle)\Big) \quad (23).$$

From **Eq. (22)** and **Eq. (23)**, although TP can try to pass the security check in **Case 1** of **Table 2** by setting $\alpha_0|e_0^i\rangle = \alpha_{15}|e_{15}^i\rangle$, it cannot obtain any information from its probe qubits using this setting of $U_e$. Hence, no unitary operation can be performed by TP in the collective attack to obtain information of the shared secret without being detected.

### 4-2 Collusion Attack

In the proposed MQSS protocol, TP can collude with some of the agents to obtain Alice's secret bits. Assume that the dishonest agents are Bob and Charlie, and they want to collaborate with TP to steal David's secret shadows. If they adopt the measure-resend attack on the qubits sent to David, their malicious actions will be detected by the security check in **Step 6** of the proposed protocol with a probability of $\frac{1}{4}$. For $m$ qubits chosen to perform the security check, the measure-resend attack will be detected with a probability of $1 - \left(\frac{3}{4}\right)^m$ where the detection rate will approach to 1 if $m$ is large enough. While the collective attack can be applied along with the collusion attack, by using similar proof in **Section 4-1**, there is no unitary operation that allows TP and the dishonest agents to obtain information of David's secret shadow. Therefore, the proposed MQSS protocol is secure against the collusion attack.

### 4-3 Trojan horse attack

Trojan horse attacks [70,71] happened pretty often in terms of implementation-dependent attacks. This paper discusses two types of Trojan horse attacks. The first type of Trojan horse attacks is by the use of a so-called delayed photon. In this attack, an attacker intercepts a qubit transmitted to a participant and then inserts a probing photon in the qubit with a delay time that is shorter than the original time window. In this method, a participant cannot detect the fake photon because it does not register on their detector. After a participant performs the corresponding operation and returns the qubit, the attacker intercepts the qubit again and separates the probing photon. In this case, the attacker can obtain full information regarding to a participant's operation by measuring the probing photon. The second type of Trojan horse attacks uses an invisible photon. The main strategy of this attack is to attach an invisible photon on each qubit sent to the participant. Since the participants cannot detect this fake photon, the attacker can steal information regarding the participant's operations in a manner similar to the use of a delay photon. That is, when the participant performs a unitary operation on the qubit, the invisible photon that attached to it will also receive the same operation simultaneously.

In the attack methods mentioned above, the attacker can only extract information regarding participant operations when they retrieve the Trojan horse photons. A two-way or circular communication protocol gives the attacker a chance to retrieve malicious photons triggered by the Trojan horse attacks. Therefore, the protocols are considered vulnerable to Trojan horse attacks if they adopt two-way or circular communication. In contrast, in a one-way communication protocol, the attacker has no chance to collect the malicious photons from the Trojan horse attacks because no qubit is returned by the participant. In other words, the protocol will be robust to Trojan horse attacks if it is a one-way communication protocol.

In the proposed MQSS protocol, the qubit transmission applies a one-way transmission strategy in which the qubits are sent only from TP to the restricted quantum participants. Although the attacker can insert probing photons into each transmitted qubit, the information of the participants' secret shadow bits cannot be obtained due to the probing photons will not be sent out after they received by the restricted quantum participants. Thus, the proposed MQSS is congenitally free from the Trojan horse attacks and all restricted quantum participants do not have to install the costly special optical devices such as the photon number splitter and the optical wavelength filter.

## 5. Conclusions

This study presents a multiparty mediated quantum secret sharing (MQSS) protocol that allows a restricted number of quantum participants to share their secret with the other restricted quantum participants via the help of a dishonest third-party (TP). Each restricted quantum participant allowed only two quantum capabilities (i.e., performing the Hadamard operation and the Z-basis measurement) and does not have to equip with any special quantum device to avoid the Trojan horse attack. In terms of quantum capabilities, the proposed MQSS is better than the existing SQSS protocols. Under the premise that a classical user owns the capability of performing the unitary operations, the proposed MQSS protocol breaks the two hard restrictions that existed in all current SQSS protocols, which are (1) the secret dealer must have full quantum capabilities, and (2) all classical quantum participants must equip with the Trojan horse detectors. There are several interesting and challenging issues can be further explored in the near future. To elaborate a few, it is worth to know that whether it is possible to improving the qubit efficiency of our MQSS protocol without any additional device and assumption. On the other hand, one can also pursue the extension the proposed MQSS protocol to a $(t, n)$- threshold QSS scheme to allow more cryptographic applications. Last but not least, the channel noises and the fault tolerance communication in quantum protocol could be another crucial issue in practice.

## Acknowledgments

We would like to thank the anonymous reviewers and the editor for their very valuable comments, which greatly enhanced the clarity of this paper. This research was partially supported by the Ministry of Science and Technology, Taiwan, R.O.C. (Grant Nos. MOST 110-2221-E-039-004, MOST 110-2221-E-143-003, MOST 110-2221-E-259-001, and MOST 110-2221-E-143-004), and China Medical University, Taiwan (Grant No. CMU110-S-21).